\documentclass[copyright,creativecommons]{eptcs}
\providecommand{\event}{PDMC 2009}
\providecommand{\volume}{~}
\providecommand{\anno}{2009}
\providecommand{\firstpage}{1}
\providecommand{\eid}{~}
\usepackage{breakurl}

\usepackage{ifthen}
\usepackage{epsfig}
\usepackage{graphics}
\usepackage{color}
\usepackage{colortbl}
\usepackage{amsmath}
\usepackage{amssymb}
\usepackage{latexsym}
\usepackage{rotating}
\usepackage{float}                
\usepackage{multirow}

\newcommand{\gfcstudent}{}  
\newcommand{\gfcvisitor}{}  

\newcommand{\id}[1]{\mathit{#1}}                 

\floatstyle{plain}                     
\newfloat{MyFloatEnv}{thb}{}[section]  
\floatname{MyFloatEnv}{Caption Name}   

\newcommand{\Naturals}{\mathbb{N}}   

\newcommand{\THEN}{\Rightarrow}                     

\newcommand{\vect}[1]{\mathbf{#1}}               
\newcommand{\matr}[1]{\mathbf{#1}}               
\newcommand{\Set}[1]{\mathcal{#1}}

\newcommand{\Top}{\id{Top}}

\newcommand{\tl}[1]{{\bf\textsf{#1}}}
\newcommand{\tU}{\tl{U}}            
\newcommand{\tE}{\tl{E}}            
\newcommand{\tEG}{\tl{EG}}          
\newcommand{\tEX}{\tl{EX}}          
\newcommand{\tEU}{\tl{EU}}          

\newcommand{\initstate}{\vs^{init}}   
\renewcommand{\initstate}{\vx_{init}}   
\newcommand{\initstateset}{\sset^{init}}  
\renewcommand{\initstateset}{\xset_{init}}  
\newcommand{\vs}{\vect{s}}            
\newcommand{\vi}{\vect{i}}            
\newcommand{\vj}{\vect{j}}            
\newcommand{\0}{\matr{0}}             
\newcommand{\1}{\matr{1}}             
\newcommand{\eset}{\Set{E}}           
\newcommand{\tset}{\Set{T}}           
\newcommand{\lset}{\Set{L}}           
\newcommand{\pset}{\Set{P}}           
\newcommand{\nset}{\Set{N}}           
\newcommand{\uset}{\Set{U}}           
\newcommand{\aset}{\Set{A}}           
\newcommand{\bset}{\Set{B}}           
\newcommand{\potxset}{\pot{\xset}}    
      
\newcommand{\vk}{\vect{k}}            
\newcommand{\wset}{\Set{W}}           
\newcommand{\xset}{\Set{X}}           
\newcommand{\yset}{\Set{Y}}           
\newcommand{\zset}{\Set{Z}}           

\newtheorem{mynote}{Note}[section]
\newtheorem{myaxiom}{Axiom}[section]
\newtheorem{mytheorem}{Theorem}[section]
\newtheorem{mycorollary}{Corollary}[section]
\newtheorem{myremark}{Remark}[section]
\newtheorem{mylemma}{Lemma}[section]
\newtheorem{mydefinition}{Definition}[section]
\newtheorem{myexample}{Example}[section]
\newtheorem{myalgorithm}{Algorithm}[section]
\newcommand{\CENTERPSSCALE}[2]{\begin{center}\mbox{\epsfig{file=#1.eps, scale=#2}}\end{center}}

\newlength{\FIRSTLISTINGSPACING} 
\newlength{\LISTINGSPACING} 
\newlength{\LISTINGINDENT} 
\newcommand{\ALGORITHM}[2]{\noindent\framebox{\small\begin{minipage}[t]{0.962\columnwidth}\vspace*{-0.0ex}\noindent \textsf{#1}\noindent\textsf{#2}\end{minipage}}}

\newcommand{\LOCAL}{\ref{}}   
\newcommand{\GLOBAL}{\ref{}}   

\newcommand{\LISTINGVARIANT}{}

\newcounter{LISTINGCOUNTER}
\newcommand{\LISTING}[2]{\vspace*{\FIRSTLISTINGSPACING}
\renewcommand{\LOCAL}{\item[] \hspace*{-1.5em}local~} 
\renewcommand{\GLOBAL}{\item[] \hspace*{-1.5em}global~} 
\begin{list}{\arabic{LISTINGCOUNTER}{\LISTINGVARIANT}}{\setlength{\labelwidth}{3em}
                                        \usecounter{LISTINGCOUNTER}}%
  {\sf \setlength{\itemsep}{\LISTINGSPACING}\setcounter{LISTINGCOUNTER}{#1}{#2}}
  \end{list}%
}
\newcommand{\LL}{\item\hspace*{0\LISTINGINDENT}}
\newcommand{\LLx}{\item\hspace*{1\LISTINGINDENT}}
\newcommand{\LLxx}{\item\hspace*{2\LISTINGINDENT}}
\newcommand{\LLxxx}{\item\hspace*{3\LISTINGINDENT}}
\newcommand{\LLxxxx}{\item\hspace*{4\LISTINGINDENT}}

\newcommand{\ASSIGN}{\mbox{$\;\leftarrow\;$}}            
\newcommand{\REMARK}[1]{\hfill {$\bullet\,${\textsl{#1}}}}  

\renewcommand{\potxset}{\xset_{pot}}    
\renewcommand{\initstate}{\vi_{init}}   

\title{
Parallel symbolic state-space exploration is difficult,\\
but what is the alternative?%
\thanks{
Work supported in part by the National Science Foundation under
grant CCF-0848463.}
}

\author{
Gianfranco Ciardo
\quad \quad
Yang Zhao
\quad \quad
Xiaoqing Jin
\institute{Department of Computer Science and Engineering\\
           University of California, Riverside}
\email{\{ciardo,zhaoy,jinx\}@cs.ucr.edu}
}

\def\titlerunning{Parallel symbolic state-space exploration is difficult}
\def\authorrunning{G.\ Ciardo, Y.\ Zhao, X.\ Jin}

\date{}
\begin{document}
\maketitle

State-space exploration is an essential first step in many modeling and
analysis problems.
Its goal is to find and store all the states reachable from the initial
state(s) of a discrete-state high-level model described, for example,
using pseudocode or Petri nets.
The state space can then be used to answer
important questions, such as ``Is there a dead state?'' and
``Can variable $n$ ever become negative?'', or as the starting point for
sophisticated investigations expressed in temporal logic.

Unfortunately, the state space is often so large that ordinary explicit
data structures and sequential algorithms simply cannot cope, prompting the
exploration of parallel or symbolic approaches.
The former uses multiple processors, from simple
networks of workstations to expensive shared-memory supercomputers or,
more recently, powerful multicore workstations, to satisfy the large
memory and run-time requirements.
The latter uses decision diagrams to compactly encode the large structured
sets and relations manipulated during state-space generation.

Both approaches have merits and limitations.
Parallel explicit state-space generation is challenging, but close
to linear speedup can be achieved, thus its scalability can be quite good;
however, the analysis is ultimately and obviously limited by the amount of
memory and number of processors available overall.
Symbolic methods rely on the heuristic properties of decision diagrams,
which can encode many, but by no means all, functions over a structured and
exponentially large domain in polynomial space;
here the pitfalls are subtler, as the performance of symbolic approaches
can vary widely depending on the particular class of decision diagram chosen,
on the order in which the variables describing the state are considered, and on
many obscure algorithmic parameters.

In this paper, we survey both approaches.
Observing that symbolic approaches
are often enormously more efficient than explicit ones
for many practical models (although it is rarely obvious \emph{a priori}
whether this will be the case on a particular application),
we argue for the need to parallelize symbolic state-space generation
algorithms, so that we can realize the advantage of both approaches.
Unfortunately, this is a very challenging endeavor, as the most
efficient symbolic algorithm, Saturation, is inherently sequential.
We conclude by discussing challenges, efforts, and promising directions
toward this goal.

\section{Introduction}

Model checking was introduced almost three decades ago and has gradually been
adopted in industrial applications.
State-space generation forms the base of safety
checking and the first step towards more complex temporal property checking.
In some scenarios, such as VLSI circuits, the potential state space is finite;
on the other hand, high-level models such as Petri Nets may have an infinite
state space.
Furthermore, even when a given Petri Net is bounded, a finite bound on
the potential state space is usually not known a priori.
Hence, state-space generation is an essential and interesting problem.

The most important metrics to evaluate the effectiveness of state-space
generation are \emph{memory consumption} and \emph{run time}.
These two metrics are often closely related, but ultimately
reflect different complexity aspects.
Considering run time, state-space generation can be expressed as
a fixpoint iteration and, for models with very large diameter
(maximum distance from an initial state to any reachable state),
this iteration can be very time consuming.
Considering memory consumption, the state space of complex models is often
too large to fit in main memory, or even in secondary storage.
Even when the latter suffices, large memory consumption leads to frequent swaps
between main and virtual memory, with negative effects on the run time.
Fortunately, processors and memory are becoming cheaper at each new generation,
so that multi-core processors and multi-processor systems provide larger
computational resources at the same or lower price.
The memory bottleneck can be relieved using more memory per workstation
and multiple workstations, but tackling the long run times is more difficult.
While multi-processor systems enable the execution of multiple
tasks in parallel, it is hard to achieve a speedup linear in the number of
processors.
The difficulty lies in the need to find enough parallel tasks to fully exploit
the available processors.
We believe that speedup will be a fundamental concern for future research on
parallel formal verification algorithms.

\subsection{Problem setting and notation}

If we ignore the particular high-level formalism used to express our system,
we are interested in studying a \emph{discrete state model} fully specified by:
\begin{itemize}
\item A set of states, or \emph{potential state space} $\potxset$,
      which describes the ``type'' of the states.
\item A set of \emph{initial states} $\initstateset \subseteq \potxset$
      from which the system behavior can evolve.
      Often, there is a single initial state, $\initstateset = \{\initstate\}$.
\item A \emph{next-state function} $\nset : \potxset \rightarrow 2^{\potxset}$,
      which describes the states to which a system can move in one step.
      This function can naturally be extended to sets of states,
      $\nset(\xset) = \bigcup_{\vi \in \xset} \nset(\vi)$.
\end{itemize}
We observe that the model is \emph{nondeterministic},
unless, for all states $\vi \in \potxset$, $|\nset(\vi)| \leq 1$, and we say
that state $\vi$ is \emph{absorbing} (or a \emph{trap}, or \emph{dead},
or a \emph{sink}) if $\nset(\vi) = \emptyset$.
In the literature, a \emph{transition relation} is sometimes defined
instead of the next-state function, but the two carry exactly
the same information: the pair of states
$(\vi,\vj)$ is in the transition relation iff $\vj \in \nset(\vi)$.
Then, we assume that the state is \emph{structured}:
\begin{itemize}
\item $\potxset = \xset_L \times \cdots \times \xset_1
      = \makebox{\Large$\times$}_{L \geq k \geq 1} \xset_k$, so that
      a (global) state is of the form $\vi = (i_L,...,i_1)$, and
      $\xset_k$ is the (discrete) \emph{local state space} for
      submodel $k$ or the \emph{local domain} for state variable $x_k$.
\end{itemize}
The techniques we consider assume that $\potxset$ is finite, thus $\xset_k$
must be finite as well, and we can map it to $\{0,1,\ldots,n_k\!-\!1\}$.
If $n_k$ is \emph{unknown a priori}, we can initially
let $\xset_k = \Naturals$, the set of natural numbers,
and discover the value of $n_k$ later, as we explore the model.
Finally, we assume \emph{asynchronous} behavior, that is,
there is a set $\eset$ of \emph{events} defining a
\emph{disjunctively-partitioned} next-state function:
\begin{itemize}
\item For each event $\alpha \!\in\! \eset$,
      $\nset_{\alpha} \!:\! \potxset \!\rightarrow\! 2^{\potxset}$.
      State $\vj$ can be reached by \emph{firing} $\alpha$ in state $\vi$
      iff $\vj \!\in\! \nset_{\alpha}(\vi)$.
\item The overall next-state function is the union of the
      functions for each event,
      $\nset(\vi) = \bigcup_{\alpha\in\eset} \nset_{\alpha}(\vi)$.
\item
We say that event $\alpha$ is \emph{enabled} in $\vi$ iff
$\nset_{\alpha}(\vi) \neq \emptyset$, otherwise we say it is \emph{disabled}.
\end{itemize}

The main goal of our study is then to generate and store the (\emph{reachable},
or \emph{actual}) \emph{state space} $\xset_{rch}$ of the model,
that is, the smallest subset of $\potxset$ containing $\initstateset$
and satisfying:
\begin{itemize}
\item The \emph{recursive definition}
      $\vi \in \xset_{rch} \wedge \vj \in \nset(\vi) \THEN \vj \in \xset_{rch}$.
\item Or, equivalently, the \emph{fixpoint equation}
      $\xset = \xset \cup \nset(\xset)$.
\end{itemize}
The most obvious way to think of the state space is as the limit of the
expression
$$
 \initstateset  \cup \nset(\initstateset) \cup \nset(\nset(\initstateset))
                \cup \nset(\nset(\nset(\initstateset))) \cup \cdots
$$
but we stress that, while this expression suggests a breadth-first
iteration where states at distance $d$ from $\initstateset$ are discovered
after exactly $d$ iterations
(i.e., after $d$ applications of the next-state functions $\nset$), this is
neither implied by the definition nor it is necessarily the most efficient
way to build $\xset_{rch}$.

Beyond state-space generation, more advanced analyses
can be performed on a discrete-state model.
For example, in the temporal logic CTL
\cite{Clarke1981CTL,McMillan1993Book},
the operators $\tEX$, $\tEU$, and $\tEG$ are \emph{complete},
that is, they can be used to express any CTL operator
through complementation, conjunction, and disjunction.
If $\aset$ and $\bset$ are the sets of states satisfying CTL
formulae $a$ and $b$, respectively, the set of states satisfying $\tEX a$ is
$\xset_{\tEX a} = \{\vi : \exists \vj \in \aset \cap \nset(\vi) \}$,
thus $\xset_{\tEX a} = \nset^{-1}(\aset)$,
where $\nset^{-1}$ is the \emph{backward state function}, i.e.,
$\nset^{-1} (\xset) = \{ \vi : \exists \vj \in \xset
                   \left(\vj \in \nset(\vi)\right)\}$.
The set of states satisfying $\tE a \tU b$ is instead
$$
\xset_{\tE a \tU b} =
\left\{
   \vi^{(0)} :
   \exists d \geq 0 \left(
      \forall c \in \{0,...,d-1\} \left(
         \vi^{(c)} \in \aset \wedge \exists \vi^{(c+1)} \in \nset(\vi^{(c)})
      \right)
      \wedge
         \vi^{(d)} \in \bset
   \right)
\right\}
$$
and can be characterized as the
\emph{smallest} solution of the fixpoint equation
$\xset \subseteq \bset \cup (\aset \cap \nset^{-1} (\xset) )$.
Analogously, the set of states satisfying $\tEG a$ is
$$
\xset_{\tEG a} =
\left\{
   \vi^{(0)} :
   \forall d \geq 0 \left(
      \forall c \in \{0,...,d-1\} \left(
         \vi^{(c)} \in \aset \wedge \exists \vi^{(c+1)} \in \nset(\vi^{(c)})
      \right)
    \right)
\right\}
$$
and can be characterized as the \emph{largest}
solution of the fixpoint equation
$\xset \supseteq \aset \cap \nset^{-1} (\xset)$.

We focus on state-space generation, but
many of the problems faced are analogous to those for the computation
of these more complex fixpoints, and many of the possible solutions
are applicable to them.

\subsection{Explicit vs. implicit techniques, which one should we parallelize?}

Symbolic model checking \cite{Burch1992} is undoubtedly a significant
breakthrough in formal verification.
Instead of representing states explicitly, symbolic approaches exploit advanced
data structures to encode and manipulate entire sets of states at once.
Paired with binary decision diagrams (BDDs) \cite{Bryant1986}, symbolic model
checkers are able to handle enormous state spaces.
At the same time, several questions remain open for BDD-based symbolic model
checking.
One of the most challenging ones is that the evolution of the BDDs being
manipulated is quite unpredictable during the fixpoint iterations, and it is
their \emph{peak size}, often many orders of
magnitude larger than the final result being sought,
that may exceed the available memory and cause the
program to fail, making symbolic algorithms brittle and subtle.
To alleviate this problem, much research has been devoted to \emph{static}
or \emph{dynamic variable ordering}, \emph{quantification scheduling},
and \emph{BDD partitioning}, to name a few.
After two decades, BDD-based symbolic techniques have become mainstream
for the verification of synchronous systems, such as VLSI circuits.

However, state-space generation for asynchronous models
such as Petri nets, communicating sequential processes, and
process algebras appears more challenging.
Although gradually replaced in synchronous systems,
explicit techniques are still competitive for asynchronous systems.
Such techniques take advantage of the locality and symmetry
properties widely enjoyed by asynchronous systems.
Partial order reduction \cite{Godefroid1996,Valmari1991CAV}
and symmetry reduction \cite{Chiola1991b}
have been successfully implemented
in explicit model checkers to reduce the number of states that must
be explored and stored, thus the run time.
These approaches explore
and store only a ``representative'' subset of the reachable states,
but are nevertheless able to answer the same questions as an
exhaustive search. Moreover, low-level memory reduction techniques
such as hash compaction are widely employed to further reduce memory
requirements.

The need to choose between explicit and implicit techniques, then, arises
mainly for the
verification and analysis of asynchronous systems, as many factors can reduce
the effectiveness of symbolic algorithms when applied to asynchronous models.
Traditional BDD techniques can efficiently encode complex synchronous transition
relations, but they are often not as compact when applied to asynchronous
events, even if disjunctive partitioning~\cite{Burch1991partitioned}
often helps.
Analogously, \emph{image computation}, the base of symbolic state-space
generation, is often expensive for asynchronous systems.
Thus, tools like SPIN~\cite{Holzmann2003spin}, a model checker
with advanced explicit techniques, have achieved wide acceptance and
success in industrial protocol verification.
Furthermore, as shown in the following sections, the parallelization
of explicit algorithms is much more successful than that of symbolic
ones, in the sense that explicit techniques can achieve almost linear
speedup when devising parallel implementations for them.

\begin{figure}
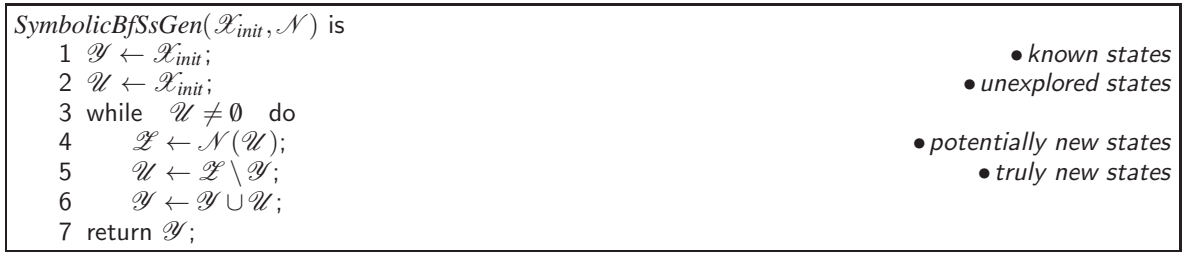

\begin{center}
\ALGORITHM{$\id{SymbolicBfSsGen}(\initstateset, \nset)$ is}{
\LISTING{0}{
\LL $\yset \ASSIGN \initstateset$; \REMARK{known states}
\LL $\uset \ASSIGN \initstateset$; \REMARK{unexplored states}
\LL while ~ $\uset \neq \emptyset$ ~ do
\LLx $\zset \ASSIGN \nset(\uset)$; \REMARK{potentially new states}
\LLx $\uset \ASSIGN \zset \setminus \yset$; \REMARK{truly new states}
\LLx $\yset \ASSIGN \yset \cup \uset$;
\LL return $\yset$;
}
}
\caption{The traditional breadth-first symbolic state-space generation.}
\label{FIG:symbolicSSgen}
\end{center}
\end{figure}

While it appears that the parallelization of explicit techniques is
more promising than that of symbolic ones, this paper argues that
symbolic techniques have nevertheless often the best chance to
shine, if not in speedup, certainly in ultimate performance.
The most traditional approach to symbolic state-space generation is
shown in Fig.~\ref{FIG:symbolicSSgen}, where all sets, i.e.,
$\yset$, $\uset$, and $\zset$, and relations, i.e., $\nset$, are
encoded using BDDs (if the local state variable $x_k$ is not
boolean, we can use $\lceil \log_2 n_k \rceil$ boolean variables for
it, or we can use MDDs, in particular \emph{extensible}
MDDs~\cite{2009SOFSEM-extensible} if $n_k$ is not known a priori; in
the following, we simply use the term DD).
This is essentially a
symbolic breadth-first exploration, where iteration $d$ discovers
all states at distance $d$ from $\initstateset$. While this approach
is often very effective, our confidence in the appropriateness of
symbolic techniques mainly comes from an even better algorithm:
\emph{Saturation}.

Initially defined for asynchronous models satisfying
\emph{Kronecker-consistency}~\cite{2001TACAS-Saturation}, where,
for each event $\alpha$,
$\nset_{\alpha}$ is the conjunction of $L$ ``local'' functions
$\nset_{\alpha,k} : \xset_k \rightarrow 2^{\xset_k}$, for $L \geq k \geq 1$,
Saturation has later been extended to a fully general setting where
each $\nset_{\alpha}$ is the conjunction of some
number of functions, each depending and affecting a subset of the state
variables~\cite{2005CHARME-GeneralSaturation}.
The main idea behind Saturation is that this disjunctive-then-conjunctive
decomposition highlights the \emph{locality} of each event $\alpha$,
so that we can define $\Top(\alpha)$ to be the highest local
state variable on which the enabling or the effect of $\alpha$ depends,
or which $\alpha$ affects.
Then, we build the DD encoding $\initstateset$ and saturate its nodes bottom up,
applying, for $k$ from $1$ to $L$, all events $\alpha$ with $\Top(\alpha) = k$
to it, exhaustively, until no more states are discovered,
with the proviso that, whenever saturating a DD node $p$ at level $k$
causes the creation of a DD node $q$ at a level below $k$, we
saturate node $q$ before completing the saturation of the higher node $p$.
Thus, when saturated, DD node $p$ encodes a fixpoint with respect to
$\nset_{\leq k} = \bigcup_{\alpha : \Top(\alpha) \leq k} \nset_{\alpha}$ and,
when we saturate the root node at level $L$, we have the entire state space,
that is, the fixpoint of $\xset = \xset \cup \nset(\xset)$.
The result is a much more efficient state-space generation algorithm
often requiring many orders of magnitude less memory and run time than
symbolic breadth-first iterations.
Indeed, we have applied the Saturation approach also to
CTL model checking~\cite{2009ATVA-ConstrainedSaturation},
distance function computation~\cite{2002FMCAD-EVMDD}, and
timed reachability in integer-timed models~\cite{2009SOFSEM-timed},
but here we will limit our discussion to state-space generation.

\subsection{Classification of previous work on parallel state-space analysis}

Modern technology offers new parallel platforms for both explicit and
implicit techniques.
In general, there are two methodologies for parallelization:
data decomposition and functional decomposition.
Data decomposition distributes the data to be processed \emph{across} parallel
tasks, each of which executes on a different workstation. 
Functional decomposition exploits the parallelism \emph{within}
a function computed by an application, allowing the distribution
of computation over multiple processors or cores.

Parallel DD-based algorithms have been developed for a variety of platforms:
shared memory multi-processor or multi-core systems \cite{Stornetta1995},
network of workstations (NOW) \cite{Milvang1998},
distributed shared memory (DSM) architectures \cite{Parasuram1994},
single-instruction-multiple-data (SIMD) and
multiple-instruction-multiple-data (MIMD) architectures \cite{Gai1995},
and vector processors \cite{Ochi1991Vector}.
According to the nature of the state-space generation algorithm,
we can generally classify the implementation platform into  two categories:
distributed-memory architecture vs.\ shared-memory architecture.
The former, e.g., NOW or PC clusters, has the advantage of possessing abundant
resources to handle large systems. 
The latter, e.g., multi-processor multi-core systems, is becoming
the predominant technology trend.
We omit detailed discussion of these platforms, and focus on their
characteristic features and challenges.
For distributed-memory systems, the main considerations are how to distribute
data, maintain load balance, and reduce communication overhead and latency.
For shared-memory systems, the mechanism employed to guarantee mutually
exclusive access to a memory region, load balance, and task scheduling
are instead paramount.
Most literature over the last 25 years has been devoted to algorithms on
distributed-memory systems due to their ability to overcome memory constraints
and to the pervasiveness of computer networks.
The advent of inexpensive memory and multi-core systems has reignited interest
in shared-memory systems.

With respect to an orthogonal classification that takes into consideration not
the hardware architecture but the type of data structure employed
by the state-space generation algorithm, two main approaches exist.
Traditional explicit state space generation approach, such as the one
used in the model-checking tool SPIN~\cite{Holzmann2003spin},
enumerates and explores each state one by one, and was first
parallelized in \cite{1998INFORMSJC-DistrGen,1997JPDC-AutomaticDistrGen,Stern97murphi}.
The other approach, DD-based state-space generation, is behind all
commonly used symbolic model-checking tools, such as
SMV \cite{McMillan1992SMV} and SMART \cite{2009PER-SMART}.

The remainder of this paper is organized as follows.
Section \ref{sec:explicit} focuses on the explicit distributed-memory
approaches proposed in \cite{BarnatASE03,BarnatBR05,Lerda99,Stern97murphi}
and the explicit shared-memory methods presented
in \cite{BarnatBR07,Barnat2008Divine,Inggs2002}.
Section \ref{sec:implicit} surveys parallel symbolic approaches
for distributed-memory architectures
\cite{Arunachalam1996,2004QEST-Distributed,2005PDMC-FirePredict,
Grumberg2005AsynPDSSGEN,Grumberg2003workefficient,Heymann2002fmsd,
Kimura1990,Stornetta1995,Stornetta1996,Yang1997},
as well as the shared-memory approach of \cite{2007CAV-Cilk}.
Section \ref{sec:challenge} discusses some promising directions for further
research on optimizing the parallelization of symbolic state-space
generation algorithms.

\section{Parallelizing explicit state-space generation}
\label{sec:explicit}

Most work in parallel explicit state-space generation has focused on
distributed-memory approaches that utilize inexpensive NOWs, although
shared memory approaches have also been explored.

\subsection{Distributed-memory approaches for explicit state-space generation}
\label{sec:Exp_Distributed}
As memory consumption is the main bottleneck for explicit techniques,
being able to exploit the overall storage available on a NOW is an appealing
idea.
Most approaches along these lines follow the general framework
of Fig.~\ref{FIG:distributedSSgen}.
Assuming that there are $N$ workstations
indexed with an identifier $w$ ranging from $1$ to $N$,
a function $\lambda : \potxset \rightarrow \{1,...,N\}$ is used to partition
the potential state space $\potxset$ into $N$ classes, that is, assign
an ``owner'' workstation to each state.
Then, each workstation performs essentially the same steps as those
required for sequential explicit state-space generation, except that each
state $\vj$ reachable from the state $\vi$ currently being explored
is checked first to see if it belongs to a different workstation,
in which case $\vj$ is sent to it, not processed locally.

\begin{figure}
\begin{center}
\ALGORITHM{
$\id{DistributedExplicitStateSpaceGeneration}(w,\initstateset,\nset,\lambda)$ is}{
\LISTING{0}{
\LLx foreach $\vi \in \initstateset$ do
\LLxx if $\lambda(\vi) = w$ then
\LLxxx $\uset_w \ASSIGN \uset_w \cup \{\vi\}$;
\LLxxx $\yset_w \ASSIGN \yset_w \cup \{\vi\}$;
\LLx repeat forever
\LLxx if $\uset_w = \emptyset$ then
\LLxxx $\uset_w \ASSIGN \id{GetStatesSentToMeFromOthers}(w)$;
       \REMARK{distributed termination detection...}
\LLxxx if $\uset_w = \emptyset$ then
       \REMARK{...returns an empty set if it is time to terminate}
\LLxxxx return $\yset_w$;
\LLxx $\vi \ASSIGN \id{ChooseOneStateFrom}(\uset_w)$;
\LLxx $\uset_w \ASSIGN \uset_w \setminus \{\vi\}$;
\LLxx foreach $\vj \in \nset(\vi)$ do
\LLxxx if $\lambda(\vj) \neq w$ then
       \REMARK{if $\vj$ does not belong to you...}
\LLxxxx $\id{SendStateToWorkstation}(\vj,\lambda(\vj))$;
        \REMARK{...send $\vj$ to its owner}
\LLxxx else if $\vj \not\in \uset_w \cup \yset_w$ then
       \REMARK{if $\vj$ belongs to you and is a new state...}
\LLxxxx $\yset_w \ASSIGN \yset_w \cup \{\vi\}$;
        \REMARK{...add it to your states...}
\LLxxxx $\uset_w \ASSIGN \uset_w \cup \{\vj\}$;
        \REMARK{...and remember to explore $\vj$ later}
}
}
\caption{General framework for distributed explicit state-space generation.}
\label{FIG:distributedSSgen}
\end{center}
\end{figure}

In this framework, several important factors affect performance.
First and foremost, the state-space partitioning function $\lambda$ has great
impact on both workload and memory balance.
Second and almost as important, the communication overhead and latency must
be taken into account when deciding how and when to exchange states.
Specifically, we need to decide whether states belonging to other workstations
are sent immediately after being discovered,
or if they are buffered into larger messages and, in this case,
how large the message buffers should be.
Then, we need to decide the frequency at which a workstation checks for
incoming states sent to it, as waiting too long to receive these states
can cause incoming buffers to grow too large, possibly with many duplicate
states in them.
The goal of tuning these parameters is then to keep all workstations busy
most of the time, while attempting to achieve similar proportions of memory
usage in each workstation and minimizing the number of message exchanges.

Of course, these factors might be in conflict.
Obviously, a partitioning function where all states belong to one workstation
has no communication at all, but the worst workload and memory balance.
More interestingly, a perfect hash function for $\lambda$ will instead
achieve excellent workload and memory balance, but also maximize
communication for state exchanges,
as the probability that any state $\vj$ reachable from $\vi$ belongs
to the same workstation as $\vi$ is only $1/N$.
Thus, a good choice of $\lambda$ should still achieve a good
workload and memory balance, but at the same time guarantee that
most state-to-state transitions remain within the same workstation,
thus require no communication.
A hash function is often used to define $\lambda$,
and an approach to achieve a good compromise was
discussed in \cite{1998INFORMSJC-DistrGen} for the case of Petri
nets, where, by hashing the state on just a few of its components (the
number of tokens in only a few of the Petri net places), we ensure that
any Petri net transition not affecting those places will result in states
belonging to the same workstation.
However, even employing this idea, it is possible to define $\lambda$
so that the mapping of \emph{reachable} states (as opposed to \emph{potential}
states) is highly uneven.

\begin{figure}
\begin{center}
\CENTERPSSCALE{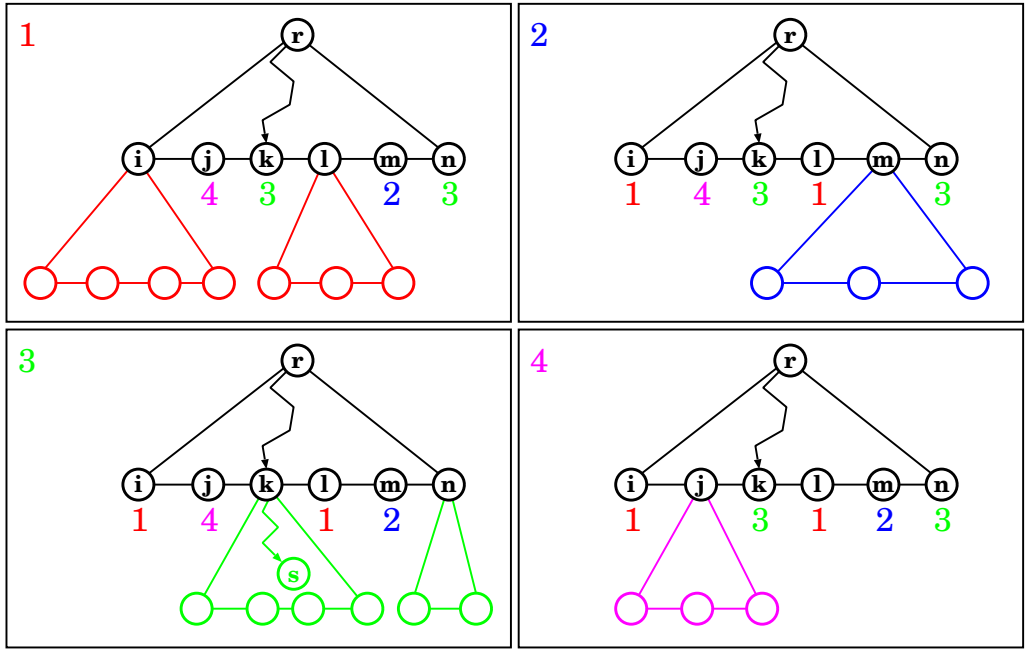}{0.65}
\end{center}
\caption{A tree-based partitioning function ($N = 4$, $M = 6$).}
\label{FIG:treepartitioning}
\end{figure}

We proposed a completely different way to define $\lambda$ in
\cite{1997JPDC-AutomaticDistrGen}, by organizing the states
in a search tree (a data structure commonly used in
explicit state-space exploration anyway), in such a way that the top
few levels of the tree are duplicated in each workstation, while the
subtrees at the lower levels are mapped to individual workstations
with no duplication. If the shared portion of the tree has $M \gg N$
terminals, each corresponding to a non-shared subtree, the approach
simply requires to associate the index of the owner workstation to
each of these terminals. As the search progresses, each workstation
keeps track of the sizes of the subtrees it owns,
and, if a workstation is overloaded, it can
restore memory balance by reassigning some of its subtrees to
light-loaded workstations.
The overhead to rebalance is clearly much
lower than with a hash function, which requires
to reallocate all the states discovered so far if it is modified.
Fig.~\ref{FIG:treepartitioning} illustrates this
idea, where the states actually stored by workstation $w$, for $ =
1,2,3,4$, are shown in the four quadrants.
When workstation $1$, $2$, or $4$ searches for state $\vs$,
with $\lambda(\vs) = 3$, it reaches the leaf storing state $\vk$ and learns
that workstation $3$ owns the subtree rooted at $\vk$.
When workstation $3$ searches for state $\vs$, it instead reaches the
node storing state $\vk$ and continues the search, until it finds $\vs$ in that
subtree, as in the figure, or determines that state $\vs$ must be
inserted in that subtree.
The shared top portion of the tree represents a small overhead in
practice, since a value for $M$ of the order of $10$ to $100$ times
$N$ is large enough in practice, and this is negligible with respect
to the number of reachable states in practical applications.
Only minimal synchronization is required, initially to agree on the
structure of the top portion of the search tree and the assignment
of its leaves to workstations, and, occasionally during state-space
generation, to agree on a different subtree-to-workstation mapping,
and broadcast this change.
The asynchronous communication between pair of workstations is then limited, as
before, to sending newly found states determined to belong to
another workstation (which again can be reduced if the top portion
of the search tree is such that the search is determined by just a
few local state components), and to exchange subtrees when load
balancing is required.

Both \cite{1998INFORMSJC-DistrGen}, which requires the user to explicitly
provide a partitioning function, and the more automated
tree-based approach of \cite{1997JPDC-AutomaticDistrGen}
achieve close to linear speedups, as well as excellent memory load balance
on conventional distributed memory architectures.

In \cite{Stern97murphi}, the explicit model checker Mur$\phi$ is
parallelized on a NOW. With the support of a fast message passing
scheme, \emph{active messages}, each process runs asynchronously
without global synchronization. A universal hash function is used to
determine the workstation to which a state belongs to and the
property of this hash function guarantees that states are
evenly distributed among the workstations.
The parallel version of Mur$\phi$ achieves close to linear speedup.
Also the tool SPIN was parallelized \cite{Lerda99}, but the focus was not on
speedup, rather on the ability to handle large models otherwise intractable.
In this work, communication becomes a dominant factor
compared to the time to compute successor states.
To minimize communication, the partition function $\lambda$ depends on one
state component.
As already shown in \cite{1998INFORMSJC-DistrGen},
this reduces cross-transitions between processes.
The parallel version of SPIN retains the most important memory and complexity
reduction techniques employed by the sequential version.

Beyond state space generation, \cite{BarnatBR05} proposed a
distributed algorithm for LTL model checking, building upon
a parallel algorithm for \emph{accepting cycle detection} in
B\"{u}chi automata \cite{BarnatASE03}.
Sequential solutions to this problem rely on depth first search (DFS),
which is hard to parallelize.
The basic idea of this work is then to detect \emph{back-level edges},
i.e., edges $(\vi,\vj)$ where the distance of state $\vi$ from the initial
state is greater than the distance of state $\vj$ from the initial state.
Parallel breadth first search (BFS) is employed to
detect back-level edges.
After each BFS step, workstations synchronize and detect back-level edges.
DFS is then employed on each workstation in parallel to find cycles.
Techniques are employed to reduce state revisiting,
and partial order reduction can be combined with this distributed algorithm.
This parallel scheme falls into the basic framework
discussed at the beginning of this section, showing that
this framework is applicable to not only reachability analysis,
but also to more complicated model checking.

In conclusion, explicit distributed-memory algorithms mainly focus on
how to maintain load balance, using static strategies such as a good
hash function or our partially-shared search tree,
and try to minimize communication by defining a partition function that
exploits knowledge of the state structure.

\subsection{Shared-memory approaches for explicit state-space generation}
\label{sec:Exp_Shared}

With respect to explicit shared-memory approaches,
\cite{BarnatBR07, Barnat2008Divine} ``port'' the parallel algorithm
from distributed-memory to shared-memory architecture in conjunction with
various techniques to improve real-world performance for LTL model checking.
First, lightweight threads play the role of workstations in the
NOW framework, and mutual exclusion techniques are used to prevent data races. 
Then, a two-level lock algorithm is used to reduce synchronization
overhead, essentially changing the lock granularity of the data structure.
Furthermore, FIFO queues handle message passing between threads, to reduce
communication overhead, and are also employed to solve memory allocation issues.
Experimental results show that the implementation scales up to 16 cores and 
has better performance than the MPI version.
\cite{BarnatBR07} concludes the main bottleneck is the state generator
and proposes in future work to balance the performance of state generator
for better scalability of the entire algorithm. 

Another relevant work is \cite{Inggs2002},
which provides another algorithm for reachability analysis
in the context of CTL* model checking.
A work stealing two-queue structure is used to dynamically maintain
load balance during state exploration, with low synchronization overhead.
Each process has an unbounded shared queuue and a bounded private queue
to store unexplored states.
A process is allowed to add and remove states in its own queues, but it can
also remove states from the shared queue of other processes, thus it
can steal another process's work instead of going idle.
Of course, shared queues must be guarded by a lock, introducing
some synchronization overhead in exchange for good load balance. 
Experimental results show almost linear speedups up to a 
12 to 16 processors, depending on the state space size;
beyond that, the benefits of using more processors are offset by
synchronization and scheduling overheads. 

This observation leads us to the second part of paper:
explicit approaches for state-space generation exhibit good scalability on
distributed and parallel systems, but only up to some point,
beyond which further increasing their performance becomes very difficult.
Since symbolic approaches tend to work very well even just in sequential
implementations, why not attempt to parallelize them?
 
\section{Parallelizing symbolic state space generation}
\label{sec:implicit}

For symbolic approaches, the first and foremost problem is to identify
a set of parallel ``tasks'', which is a challenge as most symbolic
operations are recursive and inherently sequential.
Still, there is a large amount of research attacking this problem from
different angles.
Parallelism can be realized through low-level symbolic operations, such
as logic conjunction or disjunction operations in DDs, or through high-level
algorithms, such as BFS or Saturation.
On distributed-memory systems, symbolic DD structures can be
stored in ``vertical'' or ``horizontal'' partitions, as we will
discuss in Section~\ref{sec:dis}.

\subsection{Distributed-memory approaches for symbolic state-space generation}
\label{sec:dis}

One way to achieve parallelization for symbolic techniques is designing
parallel BDD libraries for a NOW environment.
\cite{Milvang1998,Ranjan1996,Stornetta1996} discuss how to store and
manipulate BDDs on a NOW.
These approaches to achieve parallelism lie on low-level DD operations and,
if they succeed in providing an interface similar to that provided
by an ordinary sequential DD library, they can be applied to traditional
DD-based algorithms without requiring them to be rewritten.
As these packages are mostly applied to benchmark synchronous circuits,
their performance on asynchronous models is unknown.

At a higher level, parallelism can be attained through a state-space
partitioning approach similar to the one we discussed in Section
\ref{sec:Exp_Distributed}. This symbolic strategy, which we call
\emph{vertical partitioning}, assumes a partition of the potential
state space $\potxset$ into a set of $N$ ``windows'' $\{\wset_1,
\ldots , \wset_N\}$. These are in fact exactly analogous to the
function $\lambda$ required for the distributed explicit approach:
we can simply think of $\wset_w$ as $\{\vi \in \potxset : \lambda(\vi) = w\}$.
In practice, we require that the sets $\wset_w$
be easily encoded as DDs, thus they usually depend on just a few
variables (interestingly, this also bears similarity with the
requirements for a good choice of $\lambda$). The approach, shown in
Fig.~\ref{FIG:VerticalSSgen}, is similar in spirit to the explicit
one of Fig.~\ref{FIG:distributedSSgen}, except all operations are
performed symbolically on DDs, not on individual states. When
workstation $w$ explores its set of unexplored states $\uset_w$ by
applying the next-state function $\nset$, it finds both states that
it owns (states in $\wset_w$) and states that belong to other
workstations $w'$ (states in $\wset_{w'}$); the latter are sent to
the appropriate workstations, encoded as DDs.

\begin{figure}
\begin{center}
\ALGORITHM{$\id{DistributedVerticalBfsSymbolicStateSpaceGeneration}(w,\initstateset, \nset,\zset_1, ..., \zset_N)$ is}{
\vspace*{0.5ex}
\LISTING{0}{
\LL $\yset_w \ASSIGN \initstateset \cap \zset_w$; \REMARK{known states}
\LL $\uset_w \ASSIGN \initstateset \cap \zset_w$; \REMARK{unexplored states}
\LL if ~ $\uset_w = \emptyset$ ~ then
\LLx $\uset_w \ASSIGN \id{GetBddsSentToMeFromOthers}(w)$;
     \REMARK{distributed termination detection...}
\LLx if $\uset_w = \emptyset$ then
      \REMARK{...returns an empty set if it is time to terminate}
\LLxx return $\yset_w$;
\LLx $\pset_w \ASSIGN \nset(\uset_w)$; \REMARK{potentially new states}
\LLx $\uset_w \ASSIGN (\pset_w \cap \zset_w) \setminus \yset_w$;
         \REMARK{truly new states belonging to $w$}
\LLx $\yset_w \ASSIGN \yset_w \cup \uset_w$;
\LLx foreach $v \in \{1,...,w\!-\!1,w\!+\!1,...,N\}$ do
     \REMARK{distribute (potentially) new states owned by others}
\LLxx if $\pset_w \cap \zset_v \neq \emptyset$ then
\LLxxx $\id{SendBddToWorkstation}(\pset_w \cap \zset_v,v)$;
}
}
\caption{Distributed symbolic state-space generation using vertical slicing.}
\label{FIG:VerticalSSgen}
\end{center}
\end{figure}

\begin{figure}
\begin{center}
\CENTERPSSCALE{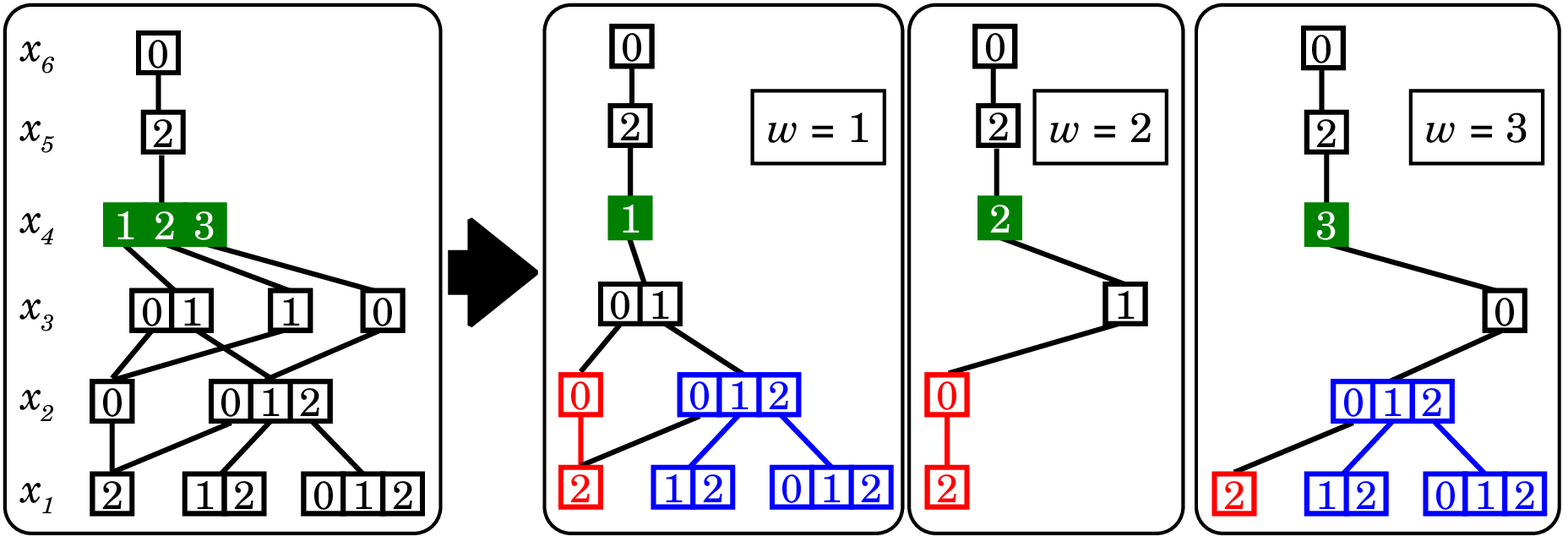}{0.65}
\CENTERPSSCALE{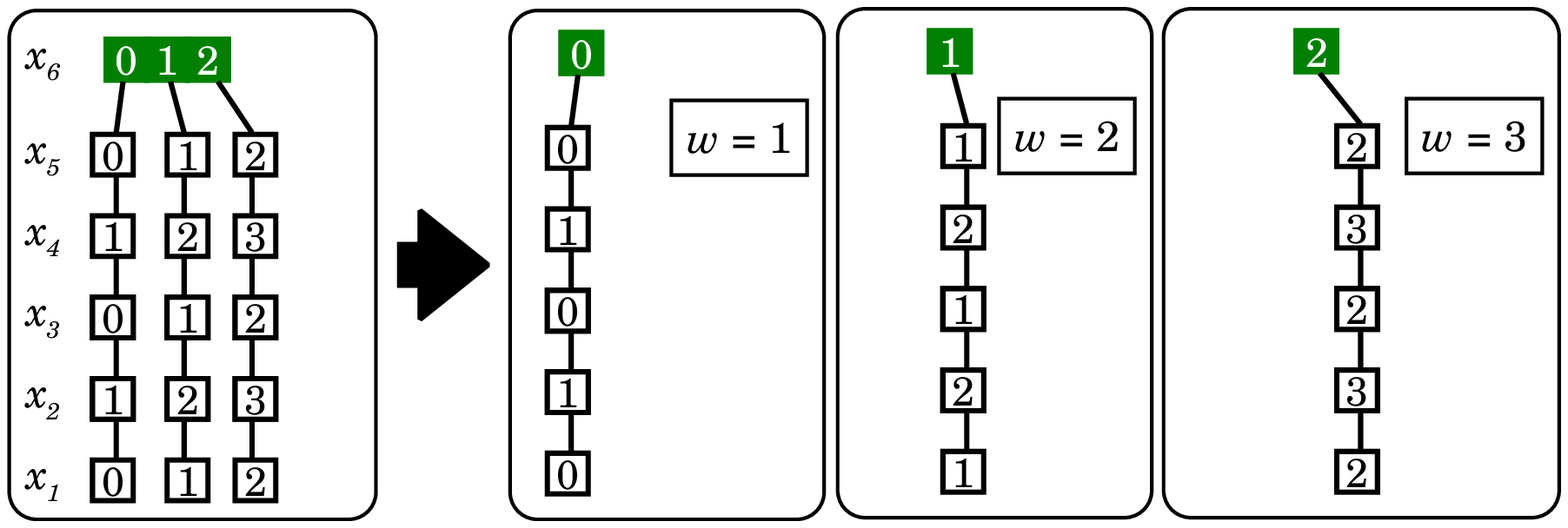}{0.65}
\end{center}
\caption{Problems with vertical partitioning.}
\label{FIG:pardis-vertical-slicing}
\end{figure}

Just as in the explicit case, the choice of partition is critical, and even
more difficult:
\begin{itemize}
\item
A balanced partition of potential states $\potxset$ does not imply
a balanced partition of reachable states $\xset_{rch}$,
yet, the slicing windows are defined on the potential state space.
\item
A balanced partition of the reachable states $\xset_{rch}$ does not imply
a balanced number of DD nodes, since
the number of states encoded by a DD is not directly related
to the number of DD nodes.
\item
Even if $\{\wset_1,...,\wset_N\}$ are a partition of potential states,
thus result in a partition of the reachable states, this does not imply absence
of DD node duplication (top of Fig.~\ref{FIG:pardis-vertical-slicing}).
Indeed, it is obvious that the minimum amount of DD node duplication
will occur when the DD is a tree, which is generally the worst case for the
application of symbolic approaches
(bottom of Fig.~\ref{FIG:pardis-vertical-slicing}).
\end{itemize}
In summary, the goal should be to minimize the sizes of the DDs managed
by the $N$ workstations (i.e., minimize the size of the largest DD, or the
the sum of the DD sizes),
but the vertical partitioning approach, being after all based on partitioning
\emph{states} and not \emph{DD nodes}, might fail to achieve this goal.

Nevertheless, some success was exhibited with this approach,
through intelligent dynamic workload balancing, which essentially reduces
to intelligent choices of windows (and reassessments of these choices).
In \cite{Heymann2002fmsd}, a slicing method is proposed to achieve balanced
slices that, with the help of a ``cost function'', also keep the number
of duplicated DD nodes low after a re-slicing.
In \cite{Grumberg2003workefficient} a more advanced strategy is
proposed to reallocate the task to free process when necessary and
release a process when its work load is small. This
\emph{work-efficient} approach attempts to adaptively minimize the
number of workstations employed at any one point during the fixed
point iterations: we start with one workstation and begin
state-space generation, then we periodically reassess the situation
and, if all workstations have an excessive memory load, we increase
$N$ and use a finer slicing, or, if they all have a very light
memory load, we decrease $N$ and use a coarser slicing. This
optimizes workstation utilization and reduces communication, so it
is a good idea (indeed, it could and should be employed also in
conjunction with the horizontal partitioning we described next).
However, this is just a confirmation that achieving true speedup
through parallelization is hard for symbolic methods.
Maximizing utilization is not the
goal, otherwise we would simply just use one workstation! In
practice, a fairly large number of workstations might be available,
whether we use them or not, so the real goal is reducing run time
for a given memory footprint, or reducing the memory footprint for a
given runtime, which is much harder.

Further improvement can be achieved using  asynchronous DD exchanges.
\cite{Grumberg2003workefficient} observed that ``the fact that the reachability
computation is synchronized in a step-by-step fashion
has a major impact on the computation time''.
Due to the need to exchange non-owned states, processes synchronize
after each image computation, and the slowest one determines the speed
of the overall computation.
To overcome this drawback, a fully asynchronous distributed algorithm is
proposed in \cite{Grumberg2005AsynPDSSGEN}, where
processes do not synchronize at each iteration, but instead run
concurrently without waiting for each other.
The classic two-phase Dijkstra algorithm is then adapted to
this framework, where the number of processes can vary dynamically.
An ``early split'' is introduced to utilize free processes and achieve speedup.
Compared to the previous approach of \cite{Grumberg2003workefficient},
improvements of up to a factor of ten are reported for some large circuits.

We now move to consider distributed approaches for the Saturation algorithm.
Saturation executes in a node-wise fashion, instead of
using heavy global breadth-first iterations, but this also means that
Saturation follows a strict order of when firing events on nodes:
a node is saturated only after all of its children have been saturated.
This policy is quite efficient but difficult to parallelize.

\begin{figure}
\begin{center}
\CENTERPSSCALE{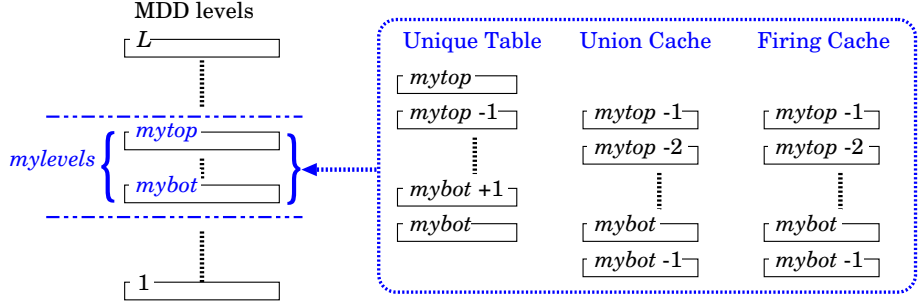}{0.74}
\end{center}
\caption{Horizontal partition used in distributed Saturation.}
\label{FIG:pardis-hor-partition}
\end{figure}

In \cite{2004QEST-Distributed} we adopted
\emph{horizontal partitioning}~\cite{Ranjan1996}
to distribute the Saturation algorithm.
Assuming that the number of levels $L$ is at least as large as the number
of workstations $N$, and hopefully much larger,
MDDs nodes are distributed to workstations according to their level:
workstation $w$ owns a contiguous range of levels
$\lset_w = \{mytop_w,...,mybot_w\}$, so that
$\{\lset_N,...,\lset_1\}$ constitute a partition of the set of levels
$\{L,...,1\}$ (see Fig.~\ref{FIG:pardis-hor-partition}).
Since Saturation fires event $\alpha$ starting at level $\Top(\alpha)$,
such an arrangement allows the appropriate workstation to start firing an
event, and, if the recursive firing reaches a boundary level,
the workstation simply issues a request to continue the operation in
the workstation responsible for the next set of levels below, and
goes idle, waiting for a reply.
The use of quasi-reduced \cite{Kimura1990} MDDs simplifies the implementation,
since it naturally allows us to associate a unique table $UT_k$
and an operation cache $OC_k$ to each level $k$.
Then, workstation $w$ stores and manages $\{UT_{mytop_w},...,UT_{mybot_w}\}$
and $\{OC_{mytop_w+1},...,OC_{mybot_w+1}\}$.
The advantage of horizontal partitioning is clear:
absolutely no duplication of nodes or cache entries.
Furthermore, memory balance simply requires to reallocate levels by changing
boundaries across neighboring workstations and moving the
corresponding nodes and cache entries, and it is easy to calculate what
the new memory load will be due to such an exchange before performing it.
However, as stated, this approach is completely sequential: at any one time,
exactly one workstation is performing work, while the
others are idle, waiting for results or to start the saturation of
nodes at their levels.
Thus, any speedup is due to being able to exploit
the overall memory of a NOW, and is observed only when comparing with
sequential Saturation running with insufficient memory.

To achieve a true speedup within this horizontal-partitioning Saturation
framework, \cite{2005PDMC-FirePredict}
employs \emph{speculative computation}, by using workstations' idle time
to perform potentially useful firings.
More precisely, we compute the relational product of a node $p$ at level $k$
in the $L$-level MDD of the current $\xset_{reach}$
and a node $r$ at level $k$ of the $2L$-level MDD encoding $\nset_{\alpha}$,
where $\Top(\alpha) > k$.
If, later on, the firing of $\alpha$ on a node at level $\Top(\alpha)$
reaches node $p$, the speculation pays off, as we simply retrieve the result
(computed using idle workstation time!) from the cache.
However, excessive speculation can easily lead to memory overload, as the
unique table and the operation cache may end up containing many useless
entries that will not be needed in future.
To reduce this problem, \cite{2006IPDPS-Speculative} associates a
\emph{firing pattern} (the set of events non-speculatively fired on a node
so far) to each MDD node, and computes a \emph{score} for each speculation
to reflect how likely the speculation of an event on a node is to move it toward
another pattern.
Furthermore, the score can take into account \emph{pattern popularity},
i.e., how many nodes have a particular pattern.
With respect to \cite{2005PDMC-FirePredict}, the results in
\cite{2006IPDPS-Speculative} show how patterns can be used to avoid excessive
speculation, so that, when speculation does not help speedup the computation,
at least it does not harm memory much.
Overall, a speedup of up to a factor of two is observed in many models
using $N = 8$ workstations, with moderate increase in memory consumption
(i.e., a workstation might use up to $1.9/8$ of the memory required
when $N = 1$, while, without speculation, the horizontal partitioning
uses only little over $1/8$ of the memory required when $N = 1$, thus
achieves almost perfect memory balance and no memory overhead, but no speedup
at all, actually a slowdown due to communication).

\subsection{Shared-memory approaches for symbolic state-space generation}

\begin{figure}
\begin{center}
\CENTERPSSCALE{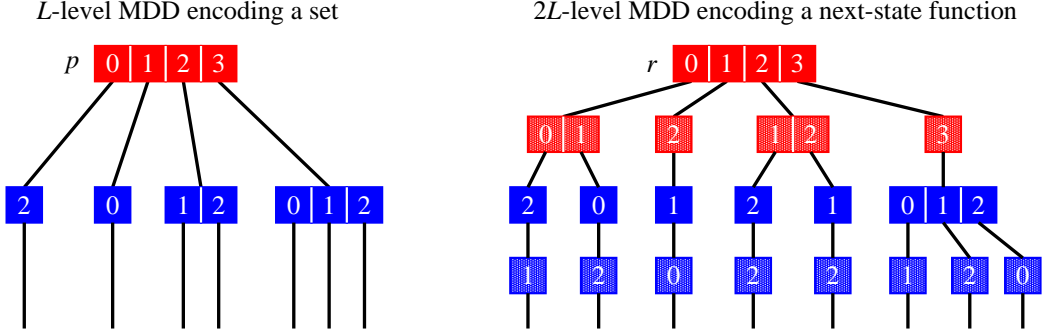}{0.74}
\end{center}
\caption{Potential fine-grained parallelism in a relational product computation.}
\label{FIG:pardis-fine-grain}
\end{figure}

With shared-memory, the main concern shifts from the
communication overhead of coarse-granularity processes to
the locking and mutual exclusion requirements of fine grained processes.
Consider a call $\id{RelProd}(p,r)$, that is, a relational product call
reaching nodes $p$ and $r$, both associated with variable $x_k$, as shown in
Fig.~\ref{FIG:pardis-fine-grain}.
The recursion will issue the calls
$\id{RelProd}(p[0],r[0][0])$,
$\id{RelProd}(p[0],r[0][1])$,
$\id{RelProd}(p[1],r[1][2])$,
$\id{RelProd}(p[2],r[2][1])$,
$\id{RelProd}(p[2],r[2][2])$, and
$\id{RelProd}(p[3],r[3][3])$.
These can be issued in any order;
indeed, they can be run in parallel, if enough cores or processors
are available.

However, DDs are not trees, they hopefully contain many
recombining paths. Thus, multiple recursive calls may reach the same
argument pairs.
In a sequential approach, we use the cache
to avoid repeating computations. In a shared-memory approach, in
addition, we need to use some locking mechanism to avoid all
redundant computations; one possible approach is for a process to
first insert its \emph{intention} to build a result in the cache, by
immediately inserting in the cache a dummy value before initiating a
computation, to be substituted by the actual result value later,
once it has been computed.
Concurrent cache lookups by other processes needing the same result
retrieve either this dummy value from the cache
(the processes will know that the value is
already being computed and will be available soon) or the actual
result (as in the sequential case).
Of course, processes must issue locks on the unique table and the operation
cache; we can avoid excessive serialization by partitioning
the unique table and cache (for example by levels),
so that the unit of memory being locked is finer, reducing blocking probability.

This works, and can indeed achieve reasonable speedup for symbolic
breadth-first state-space generation.
However, since Saturation tends to be enormously more efficient than
breadth-first iterations, we should take this approach to
parallelize Saturation. Ironically, exploring this opportunity led
us to further speedup sequential Saturation first
\cite{2006ATVA-Chaining}. One of the reasons for the efficiency of
Saturation is its extensive use of \emph{chaining}~\cite{Roig1995}:
if events $\alpha$ and $\beta$ can be fired on the set of states
$\xset$,
$$\xset \cup \nset_{\alpha}(\xset) \cup \nset_{\beta}(\xset)
   ~~ \subseteq  ~~
\xset \cup \nset_{\alpha}(\xset) \cup
\nset_{\beta}(\xset \cup \nset_{\alpha}(\xset))$$
(for the inclusion to be strict, $\alpha$ must add new states that
enable $\beta$).
Then, chaining was proposed as heuristic that looks at the system structure
(Petri net, circuit) to derive a good event order so that firing
``help compound each other's effect''.

In \cite{2006ATVA-Chaining}, we applied this idea by considering
not the structure of the high-level model, but of the MDD itself,
when performing a relational product.
Let $r_k$ be the MDD encoding
$\bigcup_{\Top(\alpha) = k} \nset_{\alpha}$.
To saturate $p$ at level $k$, we build its \emph{dynamic transition graph}
$$G_p = (\xset_k, \tset_p)  ~~~ \mbox{where} ~~~
\tset_p = \{ (i,j) \in \xset_k^2 : p[i] \neq \0 \wedge
r_k[i][j] \neq \0 \wedge p[j] \neq \1 \} .$$
If the dynamic transition graph has a path from $i$ to $j$ but not
from $j$ to $i$, then we should not issue the
call $\id{RelProd}(p[j],r_k[j][\cdot])$ until
the calls $\id{RelProd}(p[\cdot]r_k[\cdot][i])$ have converged
(i.e., they cannot add more states).
This is not a heuristic, it is \emph{guaranteed} to be optimal

\begin{figure}
\begin{center}
\CENTERPSSCALE{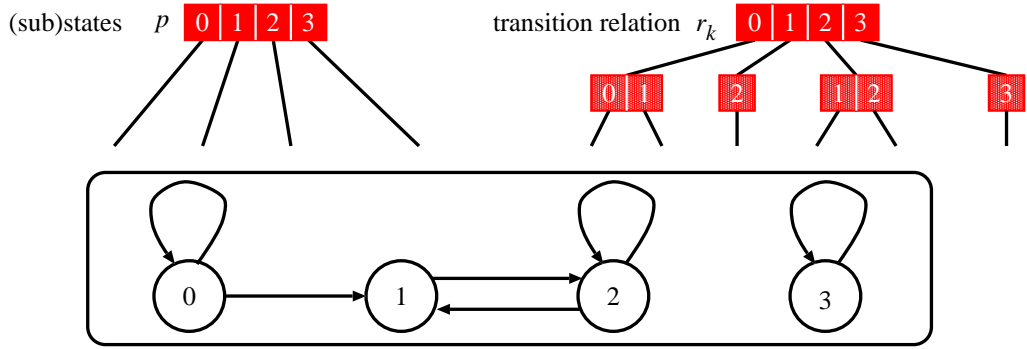}{0.74}
\end{center}
\caption{The dynamic firing graph in a relational product computation.}
\label{FIG:pardis-dynamic-graph}
\end{figure}

Unfortunately, this observation does not suffice to provide us with a total
order on the firings, since the dynamic transition graph may contain
strongly-connected components.
To ``break cycles'', we define the \emph{fullness} of node $p$ as
$\phi(p) =  \displaystyle \mbox{``number of paths encoded by $p$''} /
|\xset_k \times \cdots \times \xset_1|$, then,
under a \emph{uniform distribution assumption},
adding the result of the call $\id{RelProd}(p[i],r_k[i][j])$
increases $\phi(p[j])$ by
$$ \phi_{\Delta} \approx |\xset_k \times \cdots \times \xset_1|
   \cdot \phi(p[i]) \cdot \phi(r_k[i][j]) \cdot (1 - \phi(p[j]))$$
in expectation.
Thus, we call first $\id{RelProd}(p[i],r_k[i][j])$ for the pair $(i,j)$
maximizing $\phi_{\Delta}$.
This heuristic was shown to work quite well in practice, resulting in
consistently better run time (up to 4$\times$) and memory (up to 3$\times$)
than previous (sequential) implementations of Saturation.
Furthermore, the overhead to maintain fullness data and
to build and update $G_p$ when saturating $p$ was shown to be negligible.

Unfortunately (from a parallelization perspective), this result
improves sequential Saturation by imposing a strict order on the
$\id{RelProd}$ calls.
This fine-grained chaining heuristics
can help us understand what happens when parallelizing Saturation.
If $G_p$ contains no path from $i$ to $j$ and no path from $j$ to $i$,
we can perform some firing in parallel, out of $i$ and out of $j$, for example
(of course, we must use fine locks since the DD is a DAG, not a tree).
If $G_p$ contains a path from $i$ to $j$ but not from $j$ to $i$,
we know that we should perform the firing from $i$ on the path to $j$ before
firing any event on $j$, otherwise we may hurt chaining.
If $G_p$ contains a path from $i$ to $j$ and a path from $j$ to $i$,
we know we need to break the cycle (even in the sequential case)
and may hurt chaining, but parallelization can further hurt chaining.

This was experimentally verified in a Cilk~\cite{Blumofe1996Cilk}
implementation at York University~\cite{2007CAV-Cilk} running on a
shared-memory multicore processor computer system.
We achieved some speedup on some models on a four-core machine, but also
experienced substantial slowdowns on many models,
when parallelization hurts chaining.
Thus, this approach is likely not scalable in the number of cores
for practical models.

One intriguing possibility that needs further investigation is that we
can obviously fire in parallel on different nodes at the same level.
However, it remains to be seen how often this situation can be exploited
in practice using Saturation, especially in the common case where
$\initstateset$ contains a single state.

\section{Challenges and future goals}
\label{sec:challenge}

As we argued at the beginning of this paper, achieving good speedups
is the central goal for future work on parallel symbolic state-space
generation and formal verification.
From the above discussion, we can summarize the challenge in
the following points:
\begin{itemize}
\item Finding appropriate workload partitioning.
\item Minimizing the synchronization overhead, especially due to
      global synchronizations.
\item Devising efficient mutual exclusive schemes in DD operations.
\end{itemize}
The first point derives from the fact that DDs are brittle
in time and memory consumption during their computation, so that
balanced workloads across processes are hard to achieve and maintain.
From previous work, we can observe that distributed algorithms achieving
good speedups are mainly asynchronous, while those with global
synchronizations are often not as competitive.
On the shared-memory side,
symbolic algorithms are memory intensive, and frequent accesses to
lock-protected data can greatly reduce the potential parallelism.

We believe that the parallelization of Saturation is still a promising,
albeit challenging, work.
The main open question is how to find more tasks that
can be executed in parallel to achieve true speedup.
The reported results in \cite{2007CAV-Cilk}, which are still far from
satisfying, show that there is a subtle trade-off between the
parallelism and the level-wise firing order of Saturation.
Na\"ive parallelization of event firings will likely not lead to a faster
algorithm, as the order of firing has enormous impact on the performance.
Rather, exploring how to extract all the possible parallelism, both
at a coarse and at a fine granularity, while respecting the partial
order of operations required by the Saturation approach, is likely
to offer the greatest payback.
In addition, the speculative firing ideas used in
\cite{2005PDMC-FirePredict,2006IPDPS-Speculative,2009JLC-ImageSpeculation}
might still be helpful to provide further parallelism,
by exploiting idle processor or core time.

\bibliographystyle{eptcs}

\begin{thebibliography}{10}
\providecommand{\bibitemstart}[1]{\bibitem{#1}}
\providecommand{\bibitemend}{}
\providecommand{\bibliographystart}{}
\providecommand{\bibliographyend}{}
\providecommand{\url}[1]{\texttt{#1}}
\providecommand{\urlprefix}{Available at }
\providecommand{\bibinfo}[2]{#2}
\bibliographystart

\bibitemstart{Arunachalam1996}
\bibinfo{author}{Prakash Arunachalam}, \bibinfo{author}{Craig Chase} \&
  \bibinfo{author}{Dinos Moundanos} (\bibinfo{year}{1996}):
  \emph{\bibinfo{title}{{Distributed binary decision diagrams for verification
  of large circuits}}}.
\newblock In: {\sl \bibinfo{booktitle}{{Proc.\ Int.\ Conference on Computer
  Design (ICCD)}}}. \bibinfo{publisher}{IEEE Comp.\ Soc.\ Press},
  \bibinfo{address}{Austin, TX}, pp. \bibinfo{pages}{365--370}.
\bibitemend

\bibitemstart{BarnatASE03}
\bibinfo{author}{Jiri Barnat}, \bibinfo{author}{Lubos Brim} \&
  \bibinfo{author}{Jakub Chaloupka} (\bibinfo{year}{2003}):
  \emph{\bibinfo{title}{{Parallel breadth-first search LTL model-checking}}}.
\newblock In: {\sl \bibinfo{booktitle}{In 18th IEEE International Conference on
  Automated Software Engineering (ASE'03)}}. \bibinfo{publisher}{IEEE Comp.\
  Soc.\ Press}, pp. \bibinfo{pages}{106--115}.
\bibitemend

\bibitemstart{BarnatBR05}
\bibinfo{author}{Jiri Barnat}, \bibinfo{author}{Lubos Brim} \&
  \bibinfo{author}{Jakub Chaloupka} (\bibinfo{year}{2005}):
  \emph{\bibinfo{title}{{From distributed memory cycle detection to parallel
  LTL model checking}}}.
\newblock {\sl \bibinfo{journal}{Electronic Notes in Theoretical Computer
  Science}} \bibinfo{volume}{133}, pp. \bibinfo{pages}{21--39}.
\bibitemend

\bibitemstart{BarnatBR07}
\bibinfo{author}{Jiri Barnat}, \bibinfo{author}{Lubos Brim} \&
  \bibinfo{author}{Petr Rockai} (\bibinfo{year}{2007}):
  \emph{\bibinfo{title}{{Scalable multi-core LTL model-checking}}}.
\newblock In: {\sl \bibinfo{booktitle}{Model Checking Software, 14th
  International SPIN Workshop, Berlin, Germany, July 1-3, 2007, Proceedings}},
  {\sl \bibinfo{series}{Lecture Notes in Computer Science}}
  \bibinfo{volume}{4595}. \bibinfo{publisher}{Springer-Verlag}, pp.
  \bibinfo{pages}{187--203}.
\bibitemend

\bibitemstart{Barnat2008Divine}
\bibinfo{author}{Jiri Barnat}, \bibinfo{author}{Lubos Brim} \&
  \bibinfo{author}{Petr Ro\v{c}kai} (\bibinfo{year}{2008}):
  \emph{\bibinfo{title}{DiVinE Multi-Core --- A Parallel LTL Model-Checker}}.
\newblock In: {\sl \bibinfo{booktitle}{ATVA '08: Proceedings of the 6th
  International Symposium on Automated Technology for Verification and
  Analysis}}. \bibinfo{publisher}{Springer}, \bibinfo{address}{Seoul, Korea},
  pp. \bibinfo{pages}{234--239}.
\bibitemend

\bibitemstart{Blumofe1996Cilk}
\bibinfo{author}{R.~D. Blumofe}, \bibinfo{author}{C.~F. Joerg},
  \bibinfo{author}{B.~C. Kuszmaul}, \bibinfo{author}{C.~E. Leiserson},
  \bibinfo{author}{K.~H. Randall} \& \bibinfo{author}{Y~Zhou}
  (\bibinfo{year}{1995}): \emph{\bibinfo{title}{{Cilk: An efficient
  multithreaded runtime system}}}.
\newblock In: {\sl \bibinfo{booktitle}{{ACM SIGPLAN Symposium on Principles and
  Practice of Parallel Programming ({PPOPP}'95)}}}. \bibinfo{publisher}{ACM
  Press}, pp. \bibinfo{pages}{207--216}.
\bibitemend

\bibitemstart{Bryant1986}
\bibinfo{author}{Randy~E. Bryant} (\bibinfo{year}{1986}):
  \emph{\bibinfo{title}{{Graph-based algorithms for boolean function
  manipulation}}}.
\newblock {\sl \bibinfo{journal}{IEEE Transactions on Computers}}
  \bibinfo{volume}{35}(\bibinfo{number}{8}), pp. \bibinfo{pages}{677--691}.
\bibitemend

\bibitemstart{Burch1992}
\bibinfo{author}{J.~R. Burch}, \bibinfo{author}{E.~M. Clarke},
  \bibinfo{author}{K.~L. McMillan}, \bibinfo{author}{D.~L. Dill} \&
  \bibinfo{author}{L.~J. Hwang} (\bibinfo{year}{1992}):
  \emph{\bibinfo{title}{Symbolic model checking: $10^{20}$ states and beyond}}.
\newblock {\sl \bibinfo{journal}{Information and Computation}}
  \bibinfo{volume}{98}, pp. \bibinfo{pages}{142--170}.
\bibitemend

\bibitemstart{Burch1991partitioned}
\bibinfo{author}{Jerry~R. Burch}, \bibinfo{author}{Edmund~M. Clarke} \&
  \bibinfo{author}{David~E. Long} (\bibinfo{year}{1991}):
  \emph{\bibinfo{title}{Symbolic model checking with partitioned transition
  relations}}.
\newblock In: \bibinfo{editor}{{A. Halaas}} \& \bibinfo{editor}{{P.B. Denyer}},
  editors: {\sl \bibinfo{booktitle}{Int.\ Conference on Very Large Scale
  Integration}}. \bibinfo{organization}{IFIP Transactions},
  \bibinfo{publisher}{North-Holland}, \bibinfo{address}{Edinburgh, Scotland},
  pp. \bibinfo{pages}{49--58}.
\bibitemend

\bibitemstart{Chiola1991b}
\bibinfo{author}{G.~Chiola} \& \bibinfo{author}{Giuliana Franceschinis}
  (\bibinfo{year}{1991}): \emph{\bibinfo{title}{{A structural colour
  simplification in well-formed coloured nets}}}.
\newblock In: {\sl \bibinfo{booktitle}{{Proc.\ 4th Int.\ Workshop on Petri Nets
  and Performance Models (PNPM'91)}}}. \bibinfo{publisher}{IEEE Comp.\ Soc.\
  Press}, \bibinfo{address}{Melbourne, Australia}, pp.
  \bibinfo{pages}{144--153}.
\bibitemend

\bibitemstart{2004QEST-Distributed}
\bibinfo{author}{Ming-Ying Chung{\gfcstudent}} \& \bibinfo{author}{Gianfranco
  Ciardo} (\bibinfo{year}{2004}): \emph{\bibinfo{title}{{Saturation NOW}}}.
\newblock In: \bibinfo{editor}{Giuliana Franceschinis},
  \bibinfo{editor}{Joost-Pieter Katoen} \& \bibinfo{editor}{Murray Woodside},
  editors: {\sl \bibinfo{booktitle}{{Proc.\ Quantitative Evaluation of SysTems
  (QEST)}}}. \bibinfo{publisher}{IEEE Comp.\ Soc.\ Press},
  \bibinfo{address}{Enschede, The Netherlands}, pp. \bibinfo{pages}{272--281}.
\bibitemend

\bibitemstart{2005PDMC-FirePredict}
\bibinfo{author}{Ming-Ying Chung{\gfcstudent}} \& \bibinfo{author}{Gianfranco
  Ciardo} (\bibinfo{year}{2005}): \emph{\bibinfo{title}{{A pattern recognition
  approach for speculative firing prediction in distributed saturation
  state-space generation}}}.
\newblock In: \bibinfo{editor}{Leucker Martin} \& \bibinfo{editor}{Jaco van~de
  Pol}, editors: {\sl \bibinfo{booktitle}{Workshop on Parallel and Distributed
  Model Checking (PDMC)}}, ENTCS. \bibinfo{publisher}{Elsevier},
  \bibinfo{address}{Lisbon, Portugal}, pp. \bibinfo{pages}{65--79}.
\bibitemend

\bibitemstart{2006IPDPS-Speculative}
\bibinfo{author}{Ming-Ying Chung{\gfcstudent}} \& \bibinfo{author}{Gianfranco
  Ciardo} (\bibinfo{year}{2006}): \emph{\bibinfo{title}{{A dynamic firing
  speculation to speedup distributed symbolic state-space generation}}}.
\newblock In: \bibinfo{editor}{Arnold~L. Rosenberg}, editor: {\sl
  \bibinfo{booktitle}{Proc.\ International Parallel \& Distributed Processing
  Symposium (IPDPS)}}. \bibinfo{publisher}{IEEE Comp.\ Soc.\ Press},
  \bibinfo{address}{Rhodes, Greece}.
\newblock \bibinfo{note}{(electronic proceeding)}.
\bibitemend

\bibitemstart{2009JLC-ImageSpeculation}
\bibinfo{author}{Ming-Ying Chung{\gfcstudent}} \& \bibinfo{author}{Gianfranco
  Ciardo} (\bibinfo{year}{2009}): \emph{\bibinfo{title}{{Speculative image
  computation for distributed symbolic reachability analysis}}}.
\newblock {\sl \bibinfo{journal}{Journal of Logic and Computation}} .
\bibitemend

\bibitemstart{2006ATVA-Chaining}
\bibinfo{author}{Ming-Ying Chung{\gfcstudent}}, \bibinfo{author}{Gianfranco
  Ciardo} \& \bibinfo{author}{Andy~Jinqing Yu{\gfcstudent}}
  (\bibinfo{year}{2006}): \emph{\bibinfo{title}{{A fine-grained fullness-guided
  chaining heuristic for symbolic reachability analysis}}}.
\newblock In: \bibinfo{editor}{Susanne Graf} \& \bibinfo{editor}{Wenhui Zhang},
  editors: {\sl \bibinfo{booktitle}{{Proc.\ 4th International Symposium on
  Automated Technology for Verification and Analysis (ATVA)}}}, LNCS 4218.
  \bibinfo{publisher}{Springer-Verlag}, \bibinfo{address}{Beijing, China}, pp.
  \bibinfo{pages}{51--66}.
\bibitemend

\bibitemstart{1998INFORMSJC-DistrGen}
\bibinfo{author}{Gianfranco Ciardo}, \bibinfo{author}{Joshua Gluckman} \&
  \bibinfo{author}{David Nicol} (\bibinfo{year}{1998}):
  \emph{\bibinfo{title}{{Distributed state-space generation of discrete-state
  stochastic models}}}.
\newblock {\sl \bibinfo{journal}{INFORMS J.\ Comp.}}
  \bibinfo{volume}{10}(\bibinfo{number}{1}), pp. \bibinfo{pages}{82--93}.
\bibitemend

\bibitemstart{2001TACAS-Saturation}
\bibinfo{author}{Gianfranco Ciardo}, \bibinfo{author}{Gerald L{\"{u}}ttgen} \&
  \bibinfo{author}{Radu Siminiceanu{\gfcstudent}} (\bibinfo{year}{2001}):
  \emph{\bibinfo{title}{{Saturation: An efficient iteration strategy for
  symbolic state space generation}}}.
\newblock In: \bibinfo{editor}{Tiziana Margaria} \& \bibinfo{editor}{Wang Yi},
  editors: {\sl \bibinfo{booktitle}{{Proc.\ Tools and Algorithms for the
  Construction and Analysis of Systems (TACAS)}}}, LNCS 2031.
  \bibinfo{publisher}{Springer-Verlag}, \bibinfo{address}{Genova, Italy}, pp.
  \bibinfo{pages}{328--342}.
\bibitemend

\bibitemstart{2009PER-SMART}
\bibinfo{author}{Gianfranco Ciardo}, \bibinfo{author}{Andrew~S. Miner} \&
  \bibinfo{author}{Min Wan{\gfcstudent}} (\bibinfo{year}{2009}):
  \emph{\bibinfo{title}{{ Advanced features in SMART: the Stochastic Model
  checking Analyzer for Reliability and Timing}}}.
\newblock {\sl \bibinfo{journal}{ACM SIGMETRICS Perf.\ Eval.\ Rev.}}
  \bibinfo{volume}{36}(\bibinfo{number}{4}), pp. \bibinfo{pages}{58--63}.
\bibitemend

\bibitemstart{2002FMCAD-EVMDD}
\bibinfo{author}{Gianfranco Ciardo} \& \bibinfo{author}{Radu
  Siminiceanu{\gfcstudent}} (\bibinfo{year}{2002}):
  \emph{\bibinfo{title}{{Using edge-valued decision diagrams for symbolic
  generation of shortest paths}}}.
\newblock In: \bibinfo{editor}{Mark~D. Aagaard} \& \bibinfo{editor}{John~W.
  O'Leary}, editors: {\sl \bibinfo{booktitle}{{Proc.\ Fourth International
  Conference on Formal Methods in Computer-Aided Design (FMCAD)}}}, LNCS 2517.
  \bibinfo{publisher}{Springer-Verlag}, \bibinfo{address}{Portland, OR, USA},
  pp. \bibinfo{pages}{256--273}.
\bibitemend

\bibitemstart{2005CHARME-GeneralSaturation}
\bibinfo{author}{Gianfranco Ciardo} \& \bibinfo{author}{Andy~Jinqing
  Yu{\gfcstudent}} (\bibinfo{year}{2005}):
  \emph{\bibinfo{title}{{Saturation-based symbolic reachability analysis using
  conjunctive and disjunctive partitioning}}}.
\newblock In: \bibinfo{editor}{Dominique Borrione} \& \bibinfo{editor}{Wolfgang
  Paul}, editors: {\sl \bibinfo{booktitle}{{Proc.\ CHARME}}}, LNCS 3725.
  \bibinfo{publisher}{Springer-Verlag}, \bibinfo{address}{Saarbr{\"{u}}cken,
  Germany}, pp. \bibinfo{pages}{146--161}.
\bibitemend

\bibitemstart{Clarke1981CTL}
\bibinfo{author}{E.~M. Clarke} \& \bibinfo{author}{E.~A. Emerson}
  (\bibinfo{year}{1981}): \emph{\bibinfo{title}{{Design and synthesis of
  synchronization skeletons using branching time temporal logic}}}.
\newblock In: {\sl \bibinfo{booktitle}{{Proc.\ IBM Workshop on Logics of
  Programs}}}, LNCS 131. \bibinfo{publisher}{Springer-Verlag},
  \bibinfo{address}{London, UK}, pp. \bibinfo{pages}{52--71}.
\bibitemend

\bibitemstart{2007CAV-Cilk}
\bibinfo{author}{Jonathan Ezekiel{\gfcvisitor}}, \bibinfo{author}{Gerald
  L{\"{u}}ttgen{\gfcvisitor}} \& \bibinfo{author}{Gianfranco Ciardo}
  (\bibinfo{year}{2007}): \emph{\bibinfo{title}{{Parallelising symbolic
  state-space generators}}}.
\newblock In: \bibinfo{editor}{Werner Damm} \& \bibinfo{editor}{Holger
  Hermanns}, editors: {\sl \bibinfo{booktitle}{{Computer Aided Verification
  ({CAV}'07)}}}, LNCS 4590. \bibinfo{publisher}{Springer-Verlag},
  \bibinfo{address}{Berlin, Germany}, pp. \bibinfo{pages}{268--280}.
\bibitemend

\bibitemstart{Gai1995}
\bibinfo{author}{S.~Gai}, \bibinfo{author}{M.~Rebaudengo} \&
  \bibinfo{author}{M.~Sonza~Reorda} (\bibinfo{year}{1995}):
  \emph{\bibinfo{title}{{A data parallel algorithm for boolean function
  manipulation}}}.
\newblock In: \bibinfo{editor}{Joel Saltz} \& \bibinfo{editor}{Dennis Gannon},
  editors: {\sl \bibinfo{booktitle}{{Frontiers of Massively Parallel Scientific
  Computation (FMPSC)}}}. \bibinfo{publisher}{National Aeronautics and Space
  Administration (NASA), IEEE Comp.\ Soc.\ Press}, pp. \bibinfo{pages}{28--36}.
\bibitemend

\bibitemstart{Godefroid1996}
\bibinfo{author}{Patrice Godefroid} (\bibinfo{year}{1996}):
  \emph{\bibinfo{title}{Partial-Order Methods for the Verification of
  Concurrent Systems: An Approach to the State-Explosion Problem}}.
\newblock \bibinfo{publisher}{Springer-Verlag}.
\bibitemend

\bibitemstart{Grumberg2005AsynPDSSGEN}
\bibinfo{author}{Orna Grumberg}, \bibinfo{author}{Tamir Heyman},
  \bibinfo{author}{Nili Ifergan} \& \bibinfo{author}{Assaf Schuster}
  (\bibinfo{year}{2005}): \emph{\bibinfo{title}{Achieving Speedups in
  Distributed Symbolic Reachability Analysis Through Asynchronous
  Computation}}.
\newblock In: \bibinfo{editor}{Dominique Borrione} \& \bibinfo{editor}{Wolfgang
  Paul}, editors: {\sl \bibinfo{booktitle}{{Proc.\ Correct Hardware Design and
  Verification Methods (CHARME)}}}, {\sl \bibinfo{series}{LNCS}}
  \bibinfo{volume}{3725}. \bibinfo{publisher}{Springer-Verlag},
  \bibinfo{address}{Saarbr{\"{u}}cken, Germany}, pp. \bibinfo{pages}{129--145}.
\bibitemend

\bibitemstart{Grumberg2003workefficient}
\bibinfo{author}{Orna Grumberg}, \bibinfo{author}{Tamir Heyman} \&
  \bibinfo{author}{Assaf Schuster} (\bibinfo{year}{2003}):
  \emph{\bibinfo{title}{{A work-efficient distributed algorithm for
  reachability analysis}}}.
\newblock In: \bibinfo{editor}{Jr. Warren A.~Hunt} \& \bibinfo{editor}{Fabio
  Somenzi}, editors: {\sl \bibinfo{booktitle}{Computer Aided Verification
  (CAV)}}. \bibinfo{address}{Boulder, CO, USA}, pp. \bibinfo{pages}{54--66}.
\bibitemend

\bibitemstart{Heymann2002fmsd}
\bibinfo{author}{Tamir Heyman}, \bibinfo{author}{Danny Geist},
  \bibinfo{author}{Orna Grumberg} \& \bibinfo{author}{Assaf Schuster}
  (\bibinfo{year}{2002}): \emph{\bibinfo{title}{{A scalable parallel algorithm
  for reachability analysis of very large circuits}}}.
\newblock {\sl \bibinfo{journal}{Formal Methods in System Design}}
  \bibinfo{volume}{21}(\bibinfo{number}{3}), pp. \bibinfo{pages}{317--338}.
\bibitemend

\bibitemstart{Holzmann2003spin}
\bibinfo{author}{Gerard~J. Holzmann} (\bibinfo{year}{2003}):
  \emph{\bibinfo{title}{{The SPIN Model Checker}}}.
\newblock \bibinfo{publisher}{Addison-Wesley}.
\bibitemend

\bibitemstart{Inggs2002}
\bibinfo{author}{Cornelia~P. Inggs} \& \bibinfo{author}{Howard Barringer}
  (\bibinfo{year}{2002}): \emph{\bibinfo{title}{{Effective state exploration
  for model checking on a shared memory architecture}}}.
\newblock In: {\sl \bibinfo{booktitle}{Workshop on Parallel and Distributed
  Model Checking (PDMC'02)}}, ENTCS 68/4. \bibinfo{address}{Brno, Czech
  Republic}.
\bibitemend

\bibitemstart{McMillan1992SMV}
\bibinfo{author}{{K. L. McMillan}} (\bibinfo{year}{1992}):
  \emph{\bibinfo{title}{The {SMV} system, symbolic model checking - an
  approach}}.
\newblock \bibinfo{type}{Technical Report} \bibinfo{number}{CMU-CS-92-131},
  \bibinfo{institution}{Carnegie Mellon University}.
\bibitemend

\bibitemstart{Kimura1990}
\bibinfo{author}{Shinji Kimura} \& \bibinfo{author}{Edmund~M. Clarke}
  (\bibinfo{year}{1990}): \emph{\bibinfo{title}{{A parallel algorithm for
  constructing binary decision diagrams}}}.
\newblock In: {\sl \bibinfo{booktitle}{{Proc.\ Int.\ Conf.\ on Computer Design
  (ICCD)}}}. \bibinfo{publisher}{IEEE Comp.\ Soc.\ Press},
  \bibinfo{address}{Cambridge, MA}, pp. \bibinfo{pages}{220--223}.
\bibitemend

\bibitemstart{Lerda99}
\bibinfo{author}{Flavio Lerda} \& \bibinfo{author}{Riccardo Sisto}
  (\bibinfo{year}{1999}): \emph{\bibinfo{title}{{Distributed-memory model
  checking with SPIN}}}.
\newblock In: {\sl \bibinfo{booktitle}{Proceedings of the 5th and 6th
  International SPIN Workshops on Theoretical and Practical Aspects of SPIN
  Model Checking}}. \bibinfo{publisher}{Springer-Verlag},
  \bibinfo{address}{London, UK}, pp. \bibinfo{pages}{22--39}.
\bibitemend

\bibitemstart{McMillan1993Book}
\bibinfo{author}{K.~L. McMillan} (\bibinfo{year}{1993}):
  \emph{\bibinfo{title}{{Symbolic Model Checking}}}.
\newblock \bibinfo{publisher}{Kluwer}.
\bibitemend

\bibitemstart{Milvang1998}
\bibinfo{author}{Kim Milvang-Jensen} \& \bibinfo{author}{Alan~J. Hu}
  (\bibinfo{year}{1998}): \emph{\bibinfo{title}{{BDDNOW}: A parallel BDD
  package}}.
\newblock In: \bibinfo{editor}{Ganesh Gopalakrishnan} \&
  \bibinfo{editor}{Phillip~J. Windley}, editors: {\sl
  \bibinfo{booktitle}{Proc.\ International Confrence on Formal Methods in
  Computer-Aided Design (FMCAD)}}, {\sl \bibinfo{series}{LNCS}}
  \bibinfo{volume}{1522}. \bibinfo{publisher}{Springer-Verlag},
  \bibinfo{address}{Palo Alto, California, USA}, pp. \bibinfo{pages}{501--507}.
\bibitemend

\bibitemstart{1997JPDC-AutomaticDistrGen}
\bibinfo{author}{David Nicol} \& \bibinfo{author}{Gianfranco Ciardo}
  (\bibinfo{year}{1997}): \emph{\bibinfo{title}{{Automated parallelization of
  discrete state-space generation}}}.
\newblock {\sl \bibinfo{journal}{J.\ Par.\ and Distr.\ Comp.}}
  \bibinfo{volume}{47}, pp. \bibinfo{pages}{153--167}.
\bibitemend

\bibitemstart{Ochi1991Vector}
\bibinfo{author}{Hiroyuki Ochi}, \bibinfo{author}{Nagisa Ishiura} \&
  \bibinfo{author}{Shuzo Yajima} (\bibinfo{year}{1991}):
  \emph{\bibinfo{title}{Breadth-first manipulation of SBDD of Boolean functions
  for vector processing}}.
\newblock In: {\sl \bibinfo{booktitle}{DAC '91: Proceedings of the 28th
  ACM/IEEE Design Automation Conference}}. \bibinfo{publisher}{ACM},
  \bibinfo{address}{New York, NY, USA}, pp. \bibinfo{pages}{413--416}.
\bibitemend

\bibitemstart{Parasuram1994}
\bibinfo{author}{Y.~Parasuram}, \bibinfo{author}{E.~Stabler} \&
  \bibinfo{author}{Shiu-Kai Chin} (\bibinfo{year}{1994}):
  \emph{\bibinfo{title}{{Parallel implementation of BDD algorithms using a
  distributed shared memory}}}.
\newblock In: {\sl \bibinfo{booktitle}{{The 27th Hawaii International
  Conference on System Sciences ({HICSS}'94)}}}, ~\bibinfo{volume}{1}.
  \bibinfo{publisher}{IEEE Comp.\ Soc.\ Press}, \bibinfo{address}{Maui, HI,
  USA}, pp. \bibinfo{pages}{16--25}.
\bibitemend

\bibitemstart{Ranjan1996}
\bibinfo{author}{R.~K. Ranjan}, \bibinfo{author}{J.~V. Snaghavi},
  \bibinfo{author}{R.~K. Brayton} \&
  \bibinfo{author}{A.~Sangiovanni-Vincentelli} (\bibinfo{year}{1996}):
  \emph{\bibinfo{title}{{Binary decision diagrams on network of
  workstations}}}.
\newblock In: {\sl \bibinfo{booktitle}{{Proc.\ Int.\ Conference on Computer
  Design (ICCD)}}}. \bibinfo{publisher}{IEEE Comp.\ Soc.\ Press},
  \bibinfo{address}{Austin, TX}, pp. \bibinfo{pages}{358--364}.
\bibitemend

\bibitemstart{Roig1995}
\bibinfo{author}{O.~Roig}, \bibinfo{author}{J.~Cortadella} \&
  \bibinfo{author}{E.~Pastor} (\bibinfo{year}{1995}):
  \emph{\bibinfo{title}{{Verification of asynchronous circuits by BDD-based
  model checking of Petri nets}}}.
\newblock In: \bibinfo{editor}{Giorgio De~Michelis} \& \bibinfo{editor}{Michel
  Diaz}, editors: {\sl \bibinfo{booktitle}{{Proc.\ 16th Int.\ Conf.\ on
  Applications and Theory of Petri Nets}}}, {LNCS 935}.
  \bibinfo{publisher}{Springer-Verlag}, \bibinfo{address}{Turin, Italy}, pp.
  \bibinfo{pages}{374--391}.
\bibitemend

\bibitemstart{Stern97murphi}
\bibinfo{author}{Ulrich Stern} \& \bibinfo{author}{David~L. Dill}
  (\bibinfo{year}{1997}): \emph{\bibinfo{title}{Parallelizing the Mur$\phi$
  verifier}}.
\newblock In: \bibinfo{editor}{Orna Grumberg}, editor: {\sl
  \bibinfo{booktitle}{Proc.\ International Conference on Computer Aided
  Verification (CAV)}}, {\sl \bibinfo{series}{LNCS}} \bibinfo{volume}{1254}.
  \bibinfo{publisher}{Springer-Verlag}, \bibinfo{address}{Haifa, Israel}, pp.
  \bibinfo{pages}{256--278}.
\bibitemend

\bibitemstart{Stornetta1995}
\bibinfo{author}{Anthony~L. Stornetta} (\bibinfo{year}{1995}):
  \emph{\bibinfo{title}{{Implementation of an Efficient Parallel BDD
  Package}}}.
\newblock \bibinfo{type}{Master's thesis}, \bibinfo{school}{University of
  California, Santa Barbara}.
\bibitemend

\bibitemstart{Stornetta1996}
\bibinfo{author}{Anthony~L. Stornetta} \& \bibinfo{author}{Forrest Brewer}
  (\bibinfo{year}{1996}): \emph{\bibinfo{title}{Implementation of an efficient
  parallel {BDD} package}}.
\newblock In: {\sl \bibinfo{booktitle}{Proc.\ Design Automation Conference
  (DAC)}}. \bibinfo{publisher}{ACM Press}, \bibinfo{address}{Las Vegas, NV},
  pp. \bibinfo{pages}{641--644}.
\bibitemend

\bibitemstart{Valmari1991CAV}
\bibinfo{author}{A.~Valmari} (\bibinfo{year}{1991}): \emph{\bibinfo{title}{{A
  stubborn attack on the state explosion problem}}}.
\newblock In: {\sl \bibinfo{booktitle}{{CAV} '90}}.
  \bibinfo{publisher}{Springer-Verlag}, pp. \bibinfo{pages}{156--165}.
\bibitemend

\bibitemstart{2009SOFSEM-timed}
\bibinfo{author}{Min Wan{\gfcstudent}} \& \bibinfo{author}{Gianfranco Ciardo}
  (\bibinfo{year}{2009}): \emph{\bibinfo{title}{{Symbolic reachability analysis
  of integer timed Petri nets}}}.
\newblock In: \bibinfo{editor}{M.~Nielsen} et~al., editors: {\sl
  \bibinfo{booktitle}{{Proc.\ 35th Int.\ Conf.\ Current Trends in Theory and
  Practice of Computer Science (SOFSEM)}}}, LNCS 5404.
  \bibinfo{publisher}{Springer-Verlag},
  \bibinfo{address}{\v{S}pindler\accent23uv Ml\'{y}n, Czech Republic}, pp.
  \bibinfo{pages}{595--608}.
\bibitemend

\bibitemstart{2009SOFSEM-extensible}
\bibinfo{author}{Min Wan{\gfcstudent}} \& \bibinfo{author}{Gianfranco Ciardo}
  (\bibinfo{year}{2009}): \emph{\bibinfo{title}{{Symbolic state-space
  generation of asynchronous systems using extensible decision diagrams}}}.
\newblock In: \bibinfo{editor}{M.~Nielsen} et~al., editors: {\sl
  \bibinfo{booktitle}{{Proc.\ 35th Int.\ Conf.\ Current Trends in Theory and
  Practice of Computer Science (SOFSEM)}}}, LNCS 5404.
  \bibinfo{publisher}{Springer-Verlag},
  \bibinfo{address}{\v{S}pindler\accent23uv Ml\'{y}n, Czech Republic}, pp.
  \bibinfo{pages}{582--594}.
\bibitemend

\bibitemstart{Yang1997}
\bibinfo{author}{Bwolen Yang} \& \bibinfo{author}{David~R. O'Hallaron}
  (\bibinfo{year}{1997}): \emph{\bibinfo{title}{{Parallel breadth-first BDD
  construction}}}.
\newblock In: {\sl \bibinfo{booktitle}{{6th ACM SIGPLAN Symposium on Principles
  and Practice of Parallel Programming ({PPOPP}'97)}}}, SIGPLAN Notices 32(7).
  \bibinfo{publisher}{ACM Press}, \bibinfo{address}{Las Vegas, NV, USA}, pp.
  \bibinfo{pages}{145--156}.
\bibitemend

\bibitemstart{2009ATVA-ConstrainedSaturation}
\bibinfo{author}{Yang Zhao{\gfcstudent}} \& \bibinfo{author}{Gianfranco Ciardo}
  (\bibinfo{year}{2009}): \emph{\bibinfo{title}{{Symbolic CTL model checking of
  asynchronous systems using constrained saturation}}}.
\newblock In: \bibinfo{editor}{Zhiming Liu} \& \bibinfo{editor}{Anders~P.
  Ravn}, editors: {\sl \bibinfo{booktitle}{{Proc.\ 7th International Symposium
  on Automated Technology for Verification and Analysis (ATVA)}}}, LNCS 5799.
  \bibinfo{publisher}{Springer-Verlag}, \bibinfo{address}{Macao SAR, China},
  pp. \bibinfo{pages}{368--381}.
\bibitemend

\bibliographyend
\end{thebibliography}

\end{document}